\documentclass[aps,twocolumn,
showkeys,showpacs,nofootinbib,byrevtex]{revtex4-1}
\usepackage{amsmath,amsfonts,amssymb,amscd,amsxtra,amsthm}
\usepackage{graphicx}
\usepackage{bm}
\usepackage{ulem}
\usepackage{epsfig}

\begin{document}

\title{Features of forward $\pi N$ scattering from a Reggeized model}

\author{Kook-Jin Kong}%
\email{kong@kau.ac.kr}%

\author{Byung-Geel Yu}%
\email{bgyu@kau.ac.kr}%
\affiliation{ Research Institute of Basic Science, Korea Aerospace
University, Koyang, 10540, Korea}


\begin{abstract}
Charge exchange process $\pi^-p\to\pi^0n$ and elastic scatterings
$\pi^\pm p\to \pi^\pm p$ are investigated within the Regge
framework where the relativistic Born amplitude is reggeized for
the $t$-channel meson exchange.
Charge exchange cross section is featured by the single $\rho$ exchange.
Additional correction by Regge cuts,
$\rho$-$f_2$ and $\rho$-Pomeron, agree with differential cross sections
and new trajectory for the $\rho'(1450)$ exchange is attempted
to reproduce  polarization data.
For the description of elastic scattering data up to pion momentum
$P_{\rm Lab}\approx 250$ GeV/c, Pomeron exchange of the
Donnachie-Landshoff type is newly constructed and applied in this
work. Elastic cross section data are well reproduced with the dominance
of $f_2$ and Pomeron exchanges in intermediate and
high energies.
Analysis of nucleon resonances is presented
to test the validity of the present Regge framework below $W\leq2$ GeV.
\end{abstract}

\maketitle

\section{introduction}

A $\pi N$ system is one of the fundamental object to understand
strong interaction with its origin from QCD.
The $\pi N$ scattering near threshold offers a
testing ground for the chiral dynamics of QCD in terms of
soft pion interaction \cite{chpt}. On the other hand, the rich
structure of nucleon resonances $\Delta$ and $N^*$ in the $\pi N$
scattering below the reaction energy 2 GeV strongly supports the quark
model prediction for the baryonic spectrum  and their
properties \cite{gwu1,gwu2,kh1}. Over the resonances up to hundreds of
GeV the reaction provides information on various meson exchanges
and the nonresonant diffractive scattering that could be a manifestation
of quarks and gluon degrees of freedom rather than
hadronic degrees of freedom \cite{pichowsky}. Therefore, though
not listing a long history of theoretical development and experimental
activities initiated by $\pi N$ scattering, the reaction
should be regarded as an important source of our understanding
the dynamics of QCD in the isospin symmetry sector.

Recently, Mathieu $et\ al.$ \cite{mathieu} of JPAC
studied $\pi N$ scattering
to construct a new set of Regge amplitudes by matching the
low energy partial wave analysis with
a high energy data via the finite energy sum rule.
Nys $et\ al.$ of JPAC also analyzed world data of $KN$
charge exchange reactions
with beam energy above 5 GeV/c \cite{nys}.
Huang $et\ al.$ \cite{huang} investigated $\pi N$ charge exchange scattering
by using the Regge-cut model to provide high energy
constraints  above 2 GeV for the analysis of baryon resonances.
The primary interest of these works is in the knowledge of $\pi N$
and $KN$ scatterings at high energy in order to provide a
supplementary method for an extraction of properties of nucleon
resonances in the low energy region. However, the information
obtained from these analyses is less straightforward to current
model calculations based on the effective Lagrangian approach such
as the standard baryon pole model \cite{hung}, because the
residues in the $t$-channel helicity Regge poles fitted to empirical
data in Refs. \cite{mathieu,nys,huang} cannot communicate with coupling
strengths of hadron interactions in the Lagrangian formalism.

Therefore, it is desirable to investigate $\pi N$ scattering with
hadron models that can utilize the effective Lagrangians for the
description of the reaction beyond resonances up to the pion
momentum $P_{\rm Lab}\approx250$ GeV/c, the highest energy where
data point exists.
Unfortunately, however, there is no theory, or no model
calculations at present available for such purpose.

In this paper, we investigate $\pi N$ charge exchange and elastic
scatterings for the analysis of the reaction mechanism by the
peripheral process. For doing this, we construct the Born
amplitude to be reggeized for the meson exchange with our interest
in establishing the reaction amplitude up to such high momentum
with the interaction Lagrangians and the coupling constants
sharing with other hadron reactions.
Another issue to be addressed here is to provide the Pomeron
exchange that could be well suited for the reaction amplitude
thus constructed.
The quark-Pomeron coupling picture is introduced
to $\pi N$ elastic scattering similar to
the case of photoproductions of neutral vector mesons \cite{ pichowsky,
dl1984, laget, bgyu-phi, bgyu-omega}. Therefore, complementary to
previous findings in Refs. \cite{mathieu,nys,huang}, the result of
this work may serve to a completion of our understanding the
reaction mechanism of $\pi N$ scattering beyond
resonances.

The paper is organized as follows. In Sec. II we begin with
a statment of phenomenological features
of charge exchange scattering $\pi^- p\to\pi^0n$ to construct
the Regge amplitude for the $t$-channel $\rho$ meson exchange.
To account for the dip
in the differential cross section we introduce Regge cuts \cite{dl-cut}, and to
reproduce polarization asymmetry we consider a new trajectory
$\rho(1450)$. Sec. III
follows the similar steps as in the previous section. Features of
elastic scattering process $\pi^\pm p\to\pi^\pm p$ are introduced
and the Regge pole amplitudes relevant to these reactions are
fully constructed. The implementation of
the Pomeron exchange from the quark picture
\cite{dl1984,pichowsky,laget,titov,ysoh} is presented.
Discussion with
numerical results in experimental data at high energies
is given for total and differential
cross sections as well as polarization asymmetry.
Sec. IV devotes to an incorporation of nucleon resonances
with the Regge poles in the $t$-channel
in the $\pi N$ scattering. The Breit-Wigner
form of the nucleon resonance in the multipole
expansion of the scattering amplitude
is discussed to apply to the energy region,
below $W\leq2$ GeV.
In Section V, we
discuss our findings here and give a summary with conclusions.

\section{Charge exchange scattering}

\subsection{General features}

By charge conservation and isospin symmetry the charge exchange
reaction $\pi^-p\to\pi^0n$ allows only the single $\rho$ exchange
in the $t$-channel. Thus, it is natural to expect that
differential cross sections would show up a dip structure at the
nonsense wrong signature zero (NWSZ) of $\rho$ trajectory
$-t\approx 0.5$ (GeV/c)$^2$ from $\alpha_\rho(t)=0$. Moreover,
polarization asymmetries in this process should appear vanishing,
because the polarization asymmetry $P(\theta)$ is defined as the
interference between the spin non-flip and flip amplitudes, i.e.,
\begin{eqnarray}\label{pol-asy}
P(\theta)
=\frac{2\rm{Im}[{\cal M}^{++}{\cal M}^{+-^*}]}{|{\cal
M}^{++}|^2+|{\cal M}^{+-}|^2}\,,
\end{eqnarray}
and the single $\rho$ exchange which gives a dominant contribution
to the spin flip amplitude cannot produce nontrivial phase
interference itself between them. Therefore, the reaction needs
more theoretical consideration such as the Regge cuts and other
meson exchanges in the t-channel to reproduce the differential
cross section data and polarization asymmetry at small angles.

\subsection{Regge description}

For the reaction $\pi(k)+ N(p)\to\pi(q)+ N(p')$ process, we denote
the incoming and outgoing pion momenta by $k$ and $q$, and the
initial and final nucleon  momenta by $p$ and $p'$, respectively.
Then, conservation of 4-momentum requires $k+p=q+p'$, and
$s=(k+p)^2$, $t=(q-k)^2$, and $u=(p'-k)^2$ are the invariant
mandelstam variables. The total energy $W$ is related with the
pion momentum in the laboratory system $P_{\rm Lab}$ by the
equation $W=\sqrt{M^2+m_\pi^2+2M\sqrt{P^2_{\rm Lab}+m_\pi^2}}\ $
with $M$ and $m_\pi$ the nucleon and pion masses. The relevant
expressions for the momenta in the Lab frame and c.m. frame are
defined as
\begin{eqnarray}
&&P_{\rm
Lab}=\sqrt{\left({s-M^2-m_K^2\over2M}\right)^2-m_K^2}\,,\\
&&P_{c.m.}={1\over\sqrt{2s}}\sqrt{(s-(M+m_K)^2)(s-(M-m_K)^2)}\,,
\end{eqnarray}
respectively.

\begin{figure}[]
\centering \epsfig{file=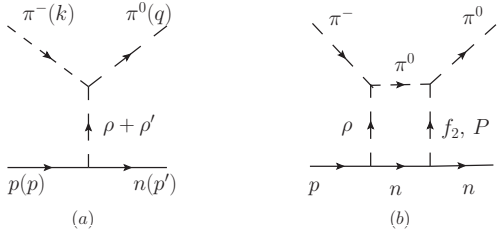, width=0.8\hsize}%
\caption{Reaction process for $\pi^- p\to \pi^0 n$  in the
$t$-channel. (a) $\rho(775)$ and $\rho'(1450)$ Regge pole
exchanges (b) Elastic cuts $\rho$-$f_2$ and $\rho$-Pomeron.}
\label{fig1}
\end{figure}

Let us begin with the scattering amplitude
simply given by the single $\rho$ exchange
\begin{eqnarray}\label{rho}
{\cal M}(\pi^- p\to \pi^0 n)=-\sqrt{2}{\cal M}_\rho\,,
\end{eqnarray}
and the Born amplitude in Fig. \ref{fig1} relevant to the $t$-channel
$\rho$ meson exchange written as
\begin{eqnarray}\label{rho-born}
{\cal M}_\rho= \Gamma^\mu_{\rho\pi\pi}(q,k)
{\Pi_{\mu\nu}^\rho(Q)\over t-m^2_\rho}\Gamma^\nu_{\rho NN}(p',p)
\,,
\end{eqnarray}
with $Q=q-k$ is the $t$-channel momentum transfer.
The coupling vertices with the spin polarization
$\Pi^{\mu\nu}_\rho$ are expressed as
\begin{eqnarray}
&&\Gamma^\mu_{\rho\pi\pi}(q,k)=g_{\rho\pi\pi}(q+k)^\mu\,,\label{rhopipi}\\
&&\Gamma^\nu_{\rho NN}(p',p)=\bar{u}(p')\biggl[g^v_{\rho
NN}\gamma^\nu+{g^t_{\rho NN}\over
4M}[\gamma^\nu,\rlap{\,/}Q]\biggr]u(p),\\
&&\Pi^{\mu\nu}_\rho(Q)=-g^{\mu\nu}+Q^\mu Q^\nu /m^2_\rho\,,
\end{eqnarray}
from the Lagrangians for  $\rho\pi\pi$ and $\rho NN$ interactions,
\begin{eqnarray}
&&{\cal
L}_{\rho\pi\pi}=g_{\rho\pi\pi}\vec\rho_\mu\cdot(\vec\pi\times\partial^\mu
\vec\pi),\\
&&{\cal L}_{\rho NN}=\bar{N}\biggl[g^v_{\rho
NN}\gamma_\nu-i{g^t_{\rho NN}\over
2M}\sigma_{\mu\nu}\partial^\nu\biggr](\tau_a)\rho^\nu N.
\end{eqnarray}

In order to describe the reaction at high energies up to tens of
GeV, we make the above Born amplitude reggeized by replacing the
Feynman propagator with the Regge one,
\begin{eqnarray}\label{reggeization}
&&{1\over t-m^2_\varphi} \rightarrow {\cal R}^{\varphi}(s,t),
\end{eqnarray}
where the Regge propagator is written as
\begin{eqnarray}\label{regge-propagator}
&&{\cal R}^{\varphi}(s,t)={\pi\alpha'_J\times{\rm
phase}\over\Gamma[\alpha_J(t)+1-J]\sin[\pi\alpha_J(t)]}
\left({s\over s_0}\right)^{\alpha_J(t)-J}
\end{eqnarray}
for the $\varphi$ meson of spin-$J$ and $s_0=1$ GeV$^2$
collectively. The trajectory of spin-$J$ meson is denoted by
$\alpha_J(t)$. The phase factor is, in general, taken to be
${1\over 2} [(-1)^J + e^{-i\pi\alpha_J(t)}]$ for the exchange
nondegenerate meson exchange.

From Eq. (\ref{rhopipi}) the decay width is evaluated as
\begin{eqnarray}\label{width}
\Gamma({\rho\to\pi\pi})={g^2_{\rho\pi\pi}k^3\over 6\pi m^2_\rho}
\end{eqnarray}
which leads to $g_{\rho\pi\pi}=5.95$ from
$\Gamma({\rho\to\pi^+\pi^-})=147.8$ MeV reported in the Particle
Data Group (PDG).  For the $\rho NN$ coupling constants, we use
$g^v_{\rho NN}=2.6$ through out this work for the
consistency with our previous works on photoproductions of
hadrons. We choose $g^t_{\rho NN}=9.62$ to be consistent with the
vector meson dominance (VMD)  for the anomalous magnetic moment
$\kappa_\rho=3.7$. However, the universality of $\rho$ meson
coupling constant is not exact between $g_{\rho\pi\pi}$ and
$2g^v_{\rho NN}$, the latter of which is estimated from the $\rho$
meson decay width $\Gamma(\rho^0\to e^+e^-)$.

\begin{figure}[]
\centering \epsfig{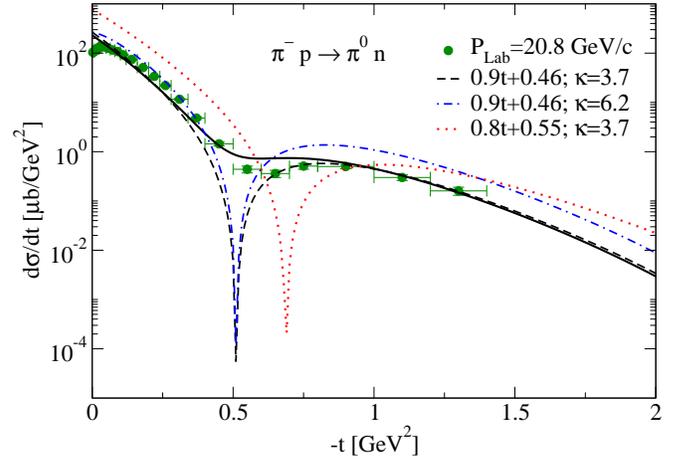}%
\caption{Differential cross section $d\sigma/dt$ for $\pi^-p \to
\pi^0 n$ scattering at $P_{\rm Lab}=20.8$ GeV/c. Cross section from the single
$\rho$ exchange with $\kappa_\rho=3.7$ and $0.9\,t+0.46$ is given
by the dashed curve, while the case with the Regge cuts,
$\rho+$$(\rho$-$f_2$+$\rho$-$\mathbb{P})$ is denoted by the solid
curve. For comparison we present cross sections from the single
$\rho$ exchange with $\kappa_\rho=3.7$ and $0.8\,t+0.55$ (red
dotted), and with $\kappa_\rho=6.2$ and $0.9\,t+0.46$ (blue
dash-dotted). Data are taken from Ref. \cite{barnes}. Additional
contribution of $\rho(1450)$ does not alter the above results.}
\label{fig2}
\end{figure}

Given the coupling constants above we take the
$\alpha_\rho(t)=0.9\,t+0.46$ from the Regge analyses of charged
$\rho$ photoproductions \cite{bgyu-rho,bgyu2} together with the
exchange nondegenerate phase for the $\rho$ Regge-pole to
reproduce the differential cross section at $P_{\rm Lab}=20.8$
GeV/c. The result is presented in Fig. \ref{fig2} by the dashed
curve.

In modelling hadron reactions involved in the $\rho$-meson
coupling to nucleon there has been
another option for $\kappa_\rho=6.2$ from the analysis of NN
Potential. Also the trajectory $\alpha_\rho=0.8\,t+0.55$ is
frequently employed with a stronger coupling constant $2g^v_{\rho
NN}\approx g_{\rho\pi\pi}$ in Regge model calculations
\cite{oh-rho}.
Without any model parameters such as cutoff masses of form factors
in this work, then, the model-dependence lies in the Regge
trajectories and meson-baryon coupling constants chosen. Thus,
by taking the advantage of the
single $\rho$ dominance of the reaction, it is
interesting to question what is the proper set for the $\rho$ meson
coupling constants and trajectory. For comparison we show
the (red) dotted curve resulting from the case of $\rho$
trajectory $0.8\,t+0.55$ with $\kappa_\rho=3.7$, and (blue)
dash-dotted one from the $0.9\,t+0.46$ with $\kappa_\rho=6.2$.  It
is clear that both cross sections are overestimating experimental
data. Moreover, the dip position at the NWSZ, $-t\approx 0.69$
(GeV/c)$^2$ in the case of $0.8\,t+0.55=0$ deviates from data.
These findings in the $\pi^-p\to\pi^0n$ support the validity of
the $\alpha_\rho=0.9\,t+0.46$ with $\kappa_\rho=3.7$ as much as in
our previous study on the charged $\rho$ photoproductions
\cite{bgyu-rho,bgyu2}.
\\

$\bullet\ $ $\rho$-cuts

As discussed above we now find a way to fill up the deep dip in
the differential cross section in Fig. \ref{fig2}.
Similar to neutral pion photoproduction $\gamma p\to\pi^0p$
\cite{dl-cut}, the $\rho$-cuts are introduced for this purpose,
and the reaction amplitude in Eq. (\ref{rho}) is now extended to
be \cite{bgyu-omega},
\begin{eqnarray}\label{rho-cut}
&&{\cal M}_\rho=\Gamma^\mu_{\rho\pi\pi}(q,k)
\Pi_{\mu\nu}^\rho(Q)\Gamma^\nu_{\rho
NN}(p',p)\nonumber\\
&&\times\left[{\cal R}^\rho(s,t)+\sum_\varphi C_\varphi
e^{d_\varphi t}e^{-i\pi\alpha^\varphi_c(t)/2}\left({s\over
s_0}\right)^{\alpha^\varphi_c(t)-1}\right],\hspace{0.5cm}
\end{eqnarray}
where $\varphi=f_2$ and $\mathbb{P}$ are the subsequent Regge pole
exchanges following the $\rho$ exchange in the elastic cut, as shown
in Fig. \ref{fig1} (b). The cut parameters $C_\varphi$ and
$d_\varphi$ represent the strength and range of the cut to be
fitted to experimental data.
The cut trajectory for
$\rho$-$\varphi$ is given by a composite of two trajectories,  i.e.,
\begin{eqnarray}
\alpha_c^\varphi={\alpha_\rho'\alpha'_\varphi\over\alpha'_\rho+\alpha'_\varphi}t
+\left(\alpha_\rho(0)+\alpha_\varphi(0)-1\right)
\end{eqnarray}
with its slope and intercept consisting of each slope and intercept.
Here we use the tensor meson $f_2$ trajectory
$\alpha_{f_2}(t)=0.9\,t+0.53$ sharing with the $\omega$
trajectory. The Pomeron trajectory is determined as
$\alpha_\mathbb{P}(t)=0.12\,t+1.06$ by the fit of elastic
scattering data at high momenta $P_{\rm Lab}=100$ and 200 GeV/c.
We shall discuss this point in Sec. III later.

In Fig. \ref{fig2} the differential cross section shows the cut
effect to fill up the dip by the $\rho$ exchange with parameters
$C_{f_2}=1.0$ GeV$^{-2}$, $d_{f_2}=2$ GeV$^{-2}$ for $\rho$-$f_2$,
and $C_\mathbb{P}=0.1$ GeV$^{-2}$, $d_\mathbb{P}=5$ GeV$^{-2}$ for
$\rho$-$\mathbb{P}$, respectively. Such an agreement with
experimental data as can be seen in Fig. \ref{fig2} appears repeatedly in
differential cross sections at other momenta at the same level of
quality.

Polarization asymmetry $P$ in the $\pi N$ scattering is the
observable that could validate the accuracy of model predictions.
As defined in Eq. (\ref{pol-asy}), it arises via the
interference between exchanges of different mesons. It is obvious that the
$\rho$ cuts could give rise to no interference, though added, because
they share the same interaction vertices with the single $\rho$ as
in Eq. (\ref{rho-cut}). Thus, we consider another isovector
exchange in the $t$-channel to find that $\rho(1450)$ of the same
quantum number $1^+(1^{--})$ with the $\rho(775)$ but of the higher
mass could be a candidate to produce nonvanishing polarization.
\\

$\bullet\ $ Daughter trajectory $\rho'(1450)$

To induce the phase interference between two different Regge
poles, we consider the daughter trajectory of $\rho$ meson of
higher mass $m_{\rho'}=1450$ MeV \cite{brodsky} with the $\pi\pi$
decay mode evident but not measured yet. (For distinction we
denote $\rho(1450)$ by $\rho'$.) Therefore, no information is
available for the $\rho'$ coupling to $\pi$, and to nucleon
either. We treat these coupling constants as parameters to fit to
polarization data. The trajectory relevant to $\rho'(1450)$ is
calculated from the relativistic quark model in Ref. \cite{ebert}
to be $\alpha_{\rho'}(t)=t-1.23$, which is different from
that of $\rho(775)$. Thus, the amplitude in Eq. (\ref{rho}) is
extended to include the amplitude ${\cal M}_{\rho'}$ of the same
form as in Eq. (\ref{rho-born}) in addition to Eq.
(\ref{rho-cut}). In the calculation we use both the $\rho$ and
$\rho'$ trajectories exchange non-degenerate. Since the intercept
of $\rho'$ trajectory is very low, the $\rho'$ exchange gives no
contribution significantly altering the differential cross section
at high momentum as in Fig. \ref{fig2}. Nevertheless, the
interference of phases between $\rho'$ and $\rho$-cuts could
reproduce the polarization data to a good degree.

\begin{figure}[]
\centering \epsfig{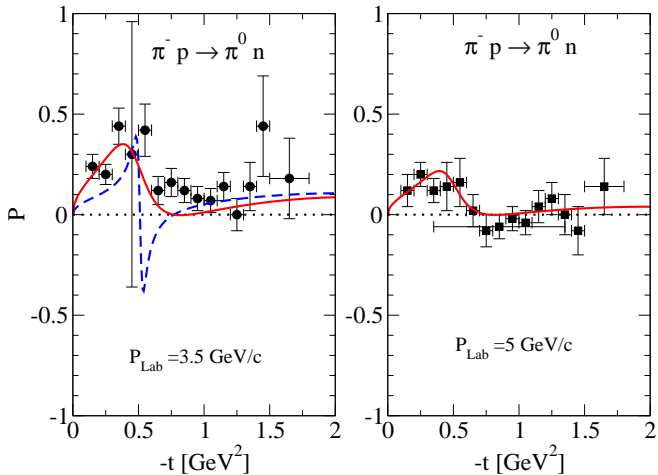}%
\caption{Polarizations for $\pi^-p \to \pi^0 n$ scattering versus
$-t$ at $P_{\rm Lab}=3.5$ and 5 GeV/c. Dotted curves are the
polarizations from $\rho$+cuts without $\rho'(1450)$. Dashed curve
is from $\rho+\rho'$ without cuts. Data are taken from Ref.
\cite{hill}.\bigskip} \label{fig3}
\end{figure}

\begin{figure}[]
\centering \epsfig{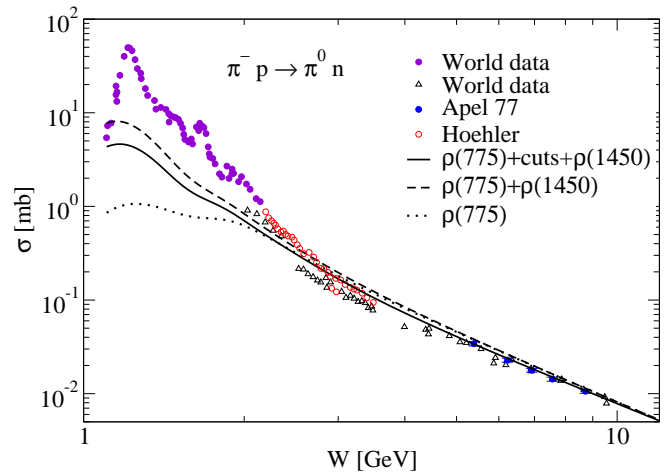}%
\caption{Total cross section for $\pi^- p\to \pi^0 n$ as a
function of invariant energy $W$. Data show the resonance peaks
below $W\approx2$ GeV. Theory and experiment coincides themselves
over the resonance region.  Data are taken from Refs.
\cite{huang,apel}. } \label{fig4}
\end{figure}

Figure \ref{fig3} shows the polarization measured at $P_{\rm
Lab}=3.5$ and 5 GeV/c in the range $0.2 \leqq -t \leqq 1.8$
(GeV/c)$^2$ where the solid curve is prediction by the full
amplitude
\begin{eqnarray}\label{full}
 -\sqrt{2}\left[\rho+\rho-{\rm
cuts}+\rho'(1450)\right].
\end{eqnarray}
At the choice of $\rho'(1450)$ coupling constants,
$G^v_{\rho'}=40$ and $G^t_{\rho'}=-75$
with $G^v_{\rho'}=g_{\rho'\pi\pi}g^v_{\rho'NN}$,
$G^t_{\rho'}=g_{\rho'\pi\pi}g^t_{\rho'NN}$,  we obtain a quite
good fit of the polarization data. Of course, adjusting these
values leads to some change of the polarization in magnitude
around $-t\approx 0.5$ (GeV/c)$^2$, but not in shape, unless the
signs of coupling constants are changed. The dotted curves are
from the single $\rho(775)$ showing null polarizations as
discussed. The dashed curve results from the $\rho(775)$ with
$\rho'(1450)$, but without $\rho$-cuts. Though dependent on the
coupling constants $G^v_{\rho'}$ and $G^t_{\rho'}$, we appreciate
highly the implication of the $\rho$-cuts and daughter $\rho'$
which result in an agreement with vanishing of the polarization at
$-t\approx 0.7$ (GeV/c)$^2$ and a slow increase as the $-t$
becomes larger.

We present total cross section in Fig. \ref{fig4} for comparison
with experimental data from threshold up to $W=10$ GeV.  The
single $\rho$ exchange is given by the dotted curve and the case
of $\rho(775)+\rho'(1450)$ by the dashed curve, respectively. The
cross section from the full amplitude $\rho+\rho'+\rho$-cuts is
depicted by the solid curve.

\section{Elastic scattering process}

\subsection{General features}

Elastic scatterings $\pi^\pm p\to \pi^\pm p$ proceed via the
$s$-channel $\Delta$ and $N^*$ excitations to show the prominent
resonance structures around $W\approx 1.2$ and 1.6 GeV,
respectively. Over the resonances the peripheral scattering of the
$t$-channel meson exchange dominates the reactions to gradually
exhibit a slow increase as the reaction energy increases, i.e.,
the diffraction scattering which is a manifestation of the Pomeron
exchange in hadron elastic reactions.
Differential cross sections have smooth $t$-dependence and no dips there.
The sign of nucleon spin polarization at the very small angle
is consistent with the sign of pion charge in the $\pi^\pm p$
reactions.

\subsection{Regge description}

Previous studies on the $\pi N$ reaction are devoted
to the analysis of total cross section which includes all the
inelastic subprocesses in addition to the elastic one
\cite{mathieu,phillips,dl-tot}. In the conventional approach where the
residues are fitted to scattering data in the $t$-channel helicity
Regge poles, the role of $\rho$ exchange was expected to account
for the difference of total cross sections between the $\pi^+p$
and $\pi^-p$ reactions in addition to the Pomeron and the second
vacuum exchange  at high energy \cite{phillips}. Recent models of
JPAC \cite{mathieu,nys} improved the Regge amplitude to include
tensor meson $f_2$ exchange  instead of the second vacuum
exchange.

In contrast to these works, however, the exclusive elastic
reactions for $\pi^\pm p\to \pi^\pm p$ are the main topics of the present section
to be investigated in the Reggeized Born term model as in previous
section. The Lagrangian formulation of hadron interactions in the
model requires those mesons that are decaying to $\pi\pi$ in the
$t$-channel exchange. In this respect exchanges of scalar meson and vector
meson $\omega$ are further included in the present work. The Pomeron
exchange is viewed from the quark-Pomeron coupling picture and we
construct a new amplitude, which is of Donnachie and Landshoff
(DL) type rather than the original version by Pichowsky
\cite{pichowsky}.

As advertised the reaction amplitude that contains all the mesons
decaying to two pions are written as,
\begin{eqnarray}
{\cal M}(\pi^\pm p)={\cal M}_\sigma\mp{\cal M}_{\omega}\mp{\cal M}_{\rho}
+{\cal M}_{f_2}+{\cal M}_{\mathbb{P}}\,,\label{eq1}
\end{eqnarray}
where the vector mesons of $C$-parity odd change sign in
accordance with pion charge. Thus, the two channels with opposite
charges $\pi^+p$ and $\pi^-p$ are distinguished by the roles of
the $\omega$ and $\rho$ meson exchange.
For the sake of consistency we share the meson-baryon coupling constants
and the Regge trajectories
in the $\pi N$ scattering with those used for meson photoproduction.
\\

$\bullet\ $ Vector mesons $\rho$ and $\omega$ exchanges

Given the $\rho$ Regge pole in the previous section, we include
the Reggeized $\omega$ exchange the same form as in Eq.
(\ref{rho-born}) with the propagator in Eq.
(\ref{regge-propagator}).  The $\omega\pi\pi$ coupling is
estimated to be $g_{\omega\pi\pi}=\pm0.18$ from the decay width
with $\Gamma({\omega\to\pi^+\pi^-})=0.13 $ MeV taken from PDG. We
use $g^v_{\omega NN}=15.6$ and $\kappa_\omega=0$ for the $\omega
NN$ couplings. The trajectory $\alpha_\omega(t)=0.9\,t+0.44$
constitutes a degenerate pair with the $f_2$ trajectory so that
both the $\omega$ and $f_2$ Regge poles share the exchange
degenerate phase in common. For the phenomenologically
better description we adopt the constant phase for both
reactions $\pi^\pm p\to \pi^\pm p$.
\\

$\bullet\ $ Scalar meson $\sigma$ exchange

The mass and full width of scalar meson $\sigma$ are reported to
be $m_\sigma=400\sim550$ MeV and $\Gamma=400\sim700$ MeV in the
PDG. The scalar meson $\sigma$ is the lightest meson to exchange.
So it could contribute to the threshold behavior of reaction cross
sections for $\pi^\pm p$ elastic scattering.

The interaction Lagrangians relevant to the coupling of $\sigma$
meson to hadrons are given by
\begin{eqnarray}
&&{\cal L}_{S}=-{1\over2}g_{\sigma\pi\pi}
m_\pi\sigma\vec\pi\cdot\vec\pi-g_{\sigma
NN}\sigma\bar{N}N\,,\\
&&{\cal L}_{V}={f_{\sigma\pi\pi}\over
2m_\pi}\sigma\partial_\mu\vec\pi\cdot
\partial^\mu\vec\pi+g_{\sigma
NN}\sigma\bar{N}N\,,
\end{eqnarray}
where the scalar and vector couplings are considered for $\sigma\pi\pi$
coupling in the manner consistent with the scalar meson $\sigma$
as the two pion $s$-wave
correlation. However, the uncertainty in the broad decay width
makes our estimate of $\sigma\pi\pi$ coupling constant very
model-dependent. A naive estimate of $g_{\sigma\pi\pi}$ by taking
$\Gamma({\sigma\to\pi\pi})=400$ MeV, for instance, from the decay
width
\begin{eqnarray}
\Gamma(\sigma\to\pi^+\pi^-)={2\over3}\Gamma(\sigma\to\pi\pi)=
{g^2_{\sigma\pi\pi} m_\pi^2 k\over12\pi m_\sigma^2},
\end{eqnarray}
yields the value $g_{\sigma\pi\pi}=\pm20.37$,
which is larger than the value $7.91\sim16.54$ extracted from the $J/\psi$
decay \cite{friesen,wu}. Here, the factor $2/3$ is
taken into account for charged channels
in the isospin space.

We now write the Born amplitude for $\sigma$ exchange as,
\begin{eqnarray}
{\cal M}_\sigma=\Gamma^{S/V}_{\sigma\pi\pi}(q,k) {1\over
t-m^2_\sigma}\Gamma_{\sigma NN}(p',p) \,,\label{sigma}
\end{eqnarray}
where the scalar and vector coupling vertices are given by
\begin{eqnarray}
&&\Gamma^{S}_{\sigma\pi\pi}(q,k)=g_{\sigma\pi\pi} m_\pi\,,\label{scpl}\\
&&\Gamma^{V}_{\sigma\pi\pi}(q,k)={f_{\sigma\pi\pi}\over
m_\pi}\,q\cdot k\,,\label{vcpl}
\end{eqnarray}
and
\begin{eqnarray}
\Gamma_{\sigma NN}(p',p)=g_{\sigma
NN}\bar{u}(p')u(p),\hspace{0.5cm} \label{eq11}
\end{eqnarray}
for the $\sigma$-meson nucleon coupling vertex.

The $\sigma$ meson exchange as the two pion correlation in
the $s$-state with a broad width was studied in the
$\pi\pi\to N\bar{N}$ reaction \cite{schutz}.
In  the $\pi N$ scattering, as a result, the $t$-channel
$\sigma$-pole derived from the dispersion relation
is expressed as
\begin{eqnarray}\label{simple}
{\cal M}_\sigma=\bar{u}(p')g_\sigma(t){t-2m_\pi^2\over m_\sigma^2-t}u(p),
\end{eqnarray}
which corresponds to
\begin{eqnarray}
&&g_\sigma(t)
={g_{\sigma NN}g_{\sigma\pi\pi}m_\pi\over 2q\cdot k}\label{scc}\,,\\
&&g_\sigma(t)
={g_{\sigma NN}f_{\sigma\pi\pi}\over 2m_\pi}\label{vcc}\,,
\end{eqnarray}
for the scalar and vector couplings in Eq. (\ref{sigma}),
respectively. These results imply that the $\sigma$-pole with
the large decay width term, $1/(t-m_\sigma^2+i\Gamma_\sigma
m_\sigma)$, in the pole model can be equivalently expressed as in
Eq. (\ref{simple}) from dispersion relation.
In Eqs. (\ref{scc}) and (\ref{vcc}), while the
latter term remains constant, the former has the energy
dependence, $1/(q\cdot k)$, which is singular at threshold.
Therefore, we favor to adopt the vector coupling scheme for
$\sigma$ exchange in Eq. (\ref{sigma}) with the coupling constant
$f_{\sigma\pi\pi}$ properly chosen to describe cross section data
near threshold. In the calculation we use $g_{\sigma NN}=14.6$
for the consistency with photoproduction of neutral vector mesons
\cite{bgyu-phi,bgyu-omega}.
\\

$\bullet\ $ Tensor meson $f_2$ exchange

For the $f_2$ tensor meson exchange, we use the following interaction Lagrangian,
\begin{eqnarray}
{\cal L}_{f_2\pi\pi}={2g_{f_2\pi\pi}\over m_{f_2}}\partial_\mu
\vec{\pi}\cdot\partial_\nu\vec{\pi}f^{\mu\nu}
\end{eqnarray}
for the $f_2\pi\pi$ coupling with  $f^{\mu\nu}$ the spin-2
tensor meson field. This gives the coupling vertex
\begin{eqnarray}
e_{\mu\nu}\Gamma^{\mu\nu}_{f_2\pi\pi}={g_{f_2\pi\pi}\over m_{f_2}}(k+q)^\mu(k+q)^\nu
\,e_{\mu\nu}
\end{eqnarray}
with $e^{\mu\nu}$ the spin-2 polarization tensor. The decay width
for $f_2\to \pi\pi$ is given by
\begin{eqnarray}
\Gamma(f_2\to \pi^+\pi^-)={2\over 3}\,\Gamma(f_2\to
\pi\pi)={4g^2_{f_2\pi\pi}\over 15\pi} {p^5\over m^4_{f_2}}
\end{eqnarray}
where $p=\sqrt{m^2_{f_2}/4-m^2_\pi}$ is the momentum of $\pi$
meson. From the full width
in the range, $121 \lesssim \Gamma({f_2})\lesssim 240$ MeV
in PDG with the branching fraction $84.2 \%$ for the
$f_2\to\pi\pi$, the coupling constant is estimated to be in the
range $4.76 \lesssim g_{f_2\pi\pi}\lesssim 6.71$
in unit of $m_{f_2}^{-1}$. 

The reaction amplitude for the $f_2$ exchange is written as
\begin{eqnarray}
{\cal M}_{f_2}= \Gamma^{\mu\nu}_{f_2\pi\pi}(q,k)
{\Pi_{\mu\nu;\alpha\beta}^{f_2}(Q)\over
t-m^2_{f_2}}\Gamma^{\alpha\beta}_{f_2 NN}(p',p) \,,\label{f2}
\end{eqnarray}
where the tensor meson-nucleon coupling vertex and the
polarization tensor for spin-2 propagation are given by
\begin{eqnarray}\label{f2cc}
&&\Gamma_{f_2NN}^{\alpha\beta}(p',p)\nonumber\\
&&=\bar{u}(p')\Bigg[{2g^{(1)}_{f_2NN}
\over M}(P^\alpha \gamma^\beta+P^\beta \gamma^\alpha)
+{4g^{(2)}_{f_2NN}\over
M^2}P^\alpha P^\beta \Bigg]u(p)\,,\nonumber\\
\end{eqnarray}
and the spin projection operator for spin-2 particle
\begin{eqnarray}
&&\Pi^{\mu\nu;\alpha\beta}_{f_2}(Q)={1\over
2}(\bar{g}^{\mu\alpha}\bar{g}^{\nu\beta}+\bar{g}^{\mu\beta}\bar{g}^{\nu\alpha})
-{1\over 3}\bar{g}^{\mu\nu}\bar{g}^{\alpha\beta}
\end{eqnarray}
with
\begin{eqnarray}
\bar{g}^{\mu\nu}=-g^{\mu\nu}+Q^\mu Q^\nu /m^2_{f_2}\,.
\end{eqnarray}

The tensor-meson nucleon  coupling constant extracted from the
tensor meson dominance was $g_{f_2NN}^{(1)}=2.12$ and
$g_{f_2NN}^{(2)}\approx0$.  But the phenomenological information
extracted from the dispersion relation as well as the partial wave
analysis for $\pi N$ scattering suggested rather the scattered
values for the $f_2 NN$ coupling constants as discussed in Ref.
\cite{oh-rho}, which showed $2.12\lesssim g^{(1)}_{f_2NN}\lesssim
7.93$ and $g^{(2)}_{f_2NN}\approx 0$.  In those meson
photoproductions involving the tensor meson exchange we used
$g_{f_2NN}^{(1)}=6.45$ and $g_{f_2NN}^{(2)}=0$ \cite{bgyu-kaon} to
agree with empirical data. For the elastic scattering of $\pi p\to
\pi p$ reaction we resume these values for the sake of
consistency, and make a list for the coupling constants and
trajectories with the corresponding phase factors in
Table~\ref{tb1}.
\\

\begin{table}[t]
\caption{Listed are the physical constants and Regge trajectories
with the corresponding phase factors for $\pi^\pm p\to\pi^\pm p$.
The symbol $\varphi$ stands for $\sigma$, $\omega$, $f_2$ and
$\rho$. For the $\sigma$ couplings $g_{\varphi\pi\pi}$  should be understood
as the vector coupling constant $f_{\sigma\pi\pi}$.
}
\begin{tabular}{c|c|c|c|c}\hline
 meson & trajectory($\alpha_\varphi$) & phase factor & $g_{\varphi\pi\pi}$ & $g^1_{\varphi NN}(g^2_{\varphi NN})$  \\
 \hline\hline
  $\rho$ &  $0.9\,t+0.46$ & $(-1+e^{-i\pi \alpha_{\rho}})/2$ & $5.95$ & $2.6$ ($9.62$)  \\%
  $\sigma$ & $0.7(t-m_{\sigma}^2)$ & $(1+e^{-i\pi \alpha_{\sigma}})/2$ & $0.5$ & 14.6 \\%
  $\omega$ &  $0.9\,t+0.44$ & $1$ & $-0.18$ & 15.6 (0)  \\%
  $f_2$ &  $0.9\,t+0.53$ & $1$ & $4.5$ & 6.45 $(0)$ \\%
\hline
 \hline
\end{tabular}\label{tb1}
\end{table}

\begin{figure}[]
\centering \epsfig{file=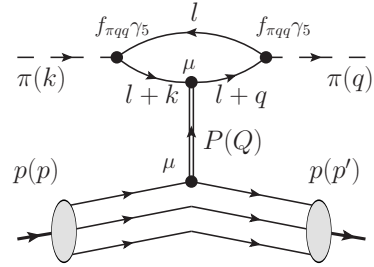, width=0.6\hsize}%
\caption{Quark diagram for the Pomeron exchange in $\pi^\pm p$
elastic scatterings. Pseudoscalar coupling $\pi\bar{q}i\gamma_5 q$
with the coupling constant $f_{\pi qq}$ is assumed at the $\pi
q\bar{q}$ coupling vertex. Momenta for quark loops are denoted by
$l$, $l+k$, and $l+q$. Quark loop of momentum $l+k$ is off
mass-shell. Vector coupling $\gamma^\mu$ is taken for the
couplings of the Pomeron-quark and Pomeron-proton currents. }
\label{fig5}
\end{figure}

$\bullet\ $ Pomeron exchange

The quark-Pomeron coupling model, as depicted in Fig. \ref{fig5},
is based on the factorization of the exclusive $\pi N$ scattering
amplitude in terms of the product of the $\pi\to q+\bar{q}\,$
fluctuation, the scattering of the $q\bar{q}$ system by the
proton, and finally the $q\bar{q}$ hadronization into a pion. From
the observation of total cross sections for $pp$, $\pi p$ and $Kp$
reactions at high energies, Donnachie and Landshoff stated that
the Pomeron couples to the separate valence quark inside a hadron
rather than to the hadron as a whole, and the strength of the
Pomeron coupling to a hadron is determined by the radius of
hadron. Therefore, assuming the quark-Pomeron coupling strength
$F_h(t)\beta_q\gamma^\mu$ with the hadron form factor $F_h(t)$ for
its size, the Pomeron contribution to the $\pi N$ cross section
can be simply written as \cite{dl1984}
\begin{eqnarray}\label{ffs}
{d\sigma\over dt}={1\over4\pi}\biggl|\left[2\beta_qF_\pi(t)\right]
(-i\alpha'_\mathbb{P}s)^{\alpha_\mathbb{P}(t)-1}\left[3\beta_{q'}F_1(t)\right]\biggr|^2
\end{eqnarray}
with the nucleon isoscalar form factor and pion form factor given
by
\begin{eqnarray}
&&
F_1(t)={4M^2-2.8t\over (4M^2-t)(1-t/0.71\ {\rm GeV}^2)^2}\,,\\
&&F_\pi(t)=\left(1-t/\Lambda^2\right)^{-n}\label{pionff}\,,
\end{eqnarray}
and the Regge-type propagator,
\begin{eqnarray}
{\cal R}^{\mathbb{P}}(s,t)=\left(\alpha'_{\mathbb
P}{s}\right)^{\alpha_{\mathbb{P}}(t)-1}e^{-i
{\pi\over2}\left[\alpha_{\mathbb{P}}(t)-1\right]}.
\end{eqnarray}
Here $\alpha_{\mathbb{P}}(t)$ is the Pomeron trajectory of the
form
\begin{eqnarray}\label{Pomeron-traj}
\alpha_\mathbb{P}(t)=\alpha'_\mathbb{P}\,t+\alpha_\mathbb{P}^0\, .
\end{eqnarray}

A more rigorous treatment of the Pomeron exchange in the $\pi N$
elastic scattering can be found in Ref. \cite{pichowsky} where the
quark-meson coupling vertices in the incoming and outgoing states
should be the Bethe-Salpeter amplitudes with the quark propagation
arising from the Dyson-Schwinger equation for the bound state of
the QCD. However, in the large momentum limit, the current quark
propagation could be replace by the free quark (constituent quark)
propagation with the constituent quark mass $m_{u(d)}\approx330$
and $m_s\approx 490$ MeV. Hence, the on-shell approximation for
the quark loops of $l$ and $l+q$ for the outgoing pion, while the
quark loop of $l+k$ considered as being off-shell with the hadron
form factor at the Pomeron-$\pi\pi$ vertex, is a good approximation
to perform the loop-integral \cite{laget}. In this work we follow
the on shell approximation as the Donnachie-Landshoff ansatz
\cite{dl} for vector meson photoproduction \cite{titov2}, and take
the pseudoscalar coupling
\begin{eqnarray}
f_{\pi qq}\bar{q}i\gamma_5 q\pi
\end{eqnarray}
for the $\pi qq$ vertex with the coupling strength $f_{\pi qq}$ in
Fig. \ref{fig5}.

The on-shell approximation leads to the loop integral simple and
the trace calculation in the loop results in the following
expression
\begin{eqnarray}
&&{\rm Tr}[(\rlap{/}l+m_q)\gamma_5((\rlap{/}l+\rlap{/}k)+m_q)
\gamma_\mu((\rlap{/}l+\rlap{/}q)+m_q)\gamma_5]\nonumber\\
&&=-4l\cdot q\,k_\mu-4l\cdot k\,q_\mu+4k\cdot q\,l_\mu,\nonumber\\
&&=m_\pi^2(k_\mu+q_\mu).\label{Pomeron2}
\end{eqnarray}
In the quark loop in Fig. \ref{fig5} the two quarks in the
outgoing pion state share the equal pion momentum, $l=-q/2$, with
the assumption that they are nearly on-shell. Then, the other
quark loop of momentum $l+k$ in the figure is off-shell and the
propagator turns out to be
\begin{eqnarray}\label{quark1}
{1\over (l+k)^2-m_q^2} ={-2\over2m_q^2-m_\pi^2/2-t}
\end{eqnarray}
with $l=-q/2$.

For the $\pi N$ elastic scattering, therefore, the Pomeron
exchange  is written as,
\begin{eqnarray}
&&{\cal M}_{\mathbb{P}}= i2F_\pi(t)\beta_q \frac{2m_\pi^2 f^2_{\pi
qq}}{2m_q^2-m_\pi^2/2-t}
F_{\mathbb{P}qq}(t)
\nonumber\\&&\hspace{1cm}\times
3F_1(t)\beta_{q'}\bar{u}(p')(\rlap{/}k+\rlap{/}q) u(p){\cal
R^\mathbb{P} }(s,t),\hspace{0.1cm}\label{pomeron}
\end{eqnarray}
where the $F_\pi(t)\beta_{q'}\gamma^\mu$ and
$F_1(t)\beta_{q'}\gamma_\mu$ with $\beta_u=2.07$ GeV$^{-1}$  and
$\beta_d=\beta_u$ are the Pomeron coupling to a quark in the pion
and in the nucleon as discussed in Eq. (\ref{ffs}). The form
factor \cite{dl}
\begin{eqnarray}
F_{\mathbb{P}qq}(t)=\frac{2\mu_0^2}{2\mu_0^2+2m_q^2-m_\pi^2/2-t}
\end{eqnarray}
is included to ensure the convergence of the off shell quark loop
with the cutoff mass $\mu_0^2=1.1$ GeV$^2$ fixed to experimental
data \cite{dl1987}.

Another quantity of new entry is the coupling constant $f_{\pi
qq}$ which is expected to obey the Goldberg-Treiman relation at
the quark level as,
\begin{eqnarray}\label{gt}
{f_{\pi qq}\over 2m_q}={1\over2f_\pi}{3\over5}g_A\ .
\end{eqnarray}
Given the nucleon axial charge  $g_A=1.25$,  pion decay constant
$f_\pi$=93.1 MeV, and by using the quark mass $m_q=330$ MeV we
determine $f_{\pi qq}=2.65$.

It is worth noting in Eq. (\ref{quark1}) that the
$\left(t/2+k^2/2-q^2/4-m_q^2\right)^{-1}$ becomes singular near $-t\approx 0$ as
$k^2$=$q^2$=$m^2_\pi$ for the pion elastic scattering when
$m_\pi=2m_q$ is assumed.
For a better convergence of the quark loop in addition to the form
factor $F_{\mathbb{P}qq}(t)$, therefore, we utilize the pion form
factor $F_\pi(t)$. Moreover, in order to adjust the range of the
$F_\pi(t)$ it is convenient to use the cutoff mass in Eq.
(\ref{pionff}) having an energy dependence as
\begin{eqnarray}\label{lambdamass}
\Lambda(k)={k\over \mu}(W-W_{th}),
\end{eqnarray}
where $k$ is the incident pion momentum in the c.m. system, $\mu$
is the parameter of mass unit, and $W_{th}$ is the total energy at threshold.

\begin{figure}[]
\centering \epsfig{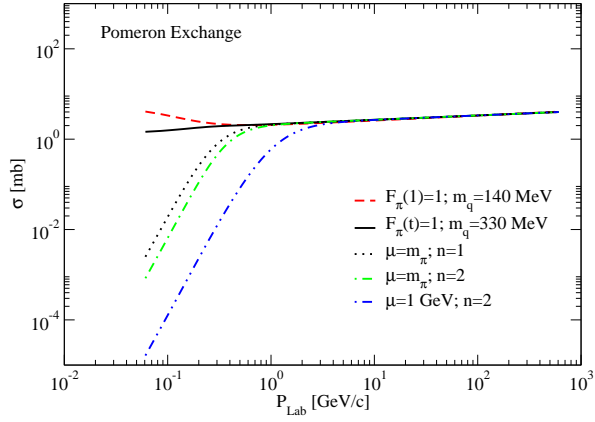}%
\caption{Dependence of Pomeron exchange on $F_\pi(t)$. Given the
physical pion mass $m_\pi$ and Pomeron trajectory in Eq.
(\ref{pomeron-traj}), the red dashed curve shows the divergence
for the $m_q=140$ MeV in the absence of $F_\pi(t)$. $f_{\pi
qq}=1.32$ is taken for a coincident with others for comparison.
The rest of
curves are resulting from the change of mass parameter $\mu$ and
power $n$ with $m_q=330$ MeV fixed. } \label{fig6}
\end{figure}

Figure \ref{fig6} shows the divergence of the Pomeron exchange
depending on the quark mass, for instance, $m_q=140$ MeV taken
without the form factor $F_\pi(t)$. Dotted, dash-dotted, and
dash-dot-dotted are the cases of Pomeron converging in the lower
energy region due to the role of $F_\pi(t)$ with the parameter
$\mu$ and power $n$ as designated in the figure.

\begin{figure}[]
\centering \epsfig{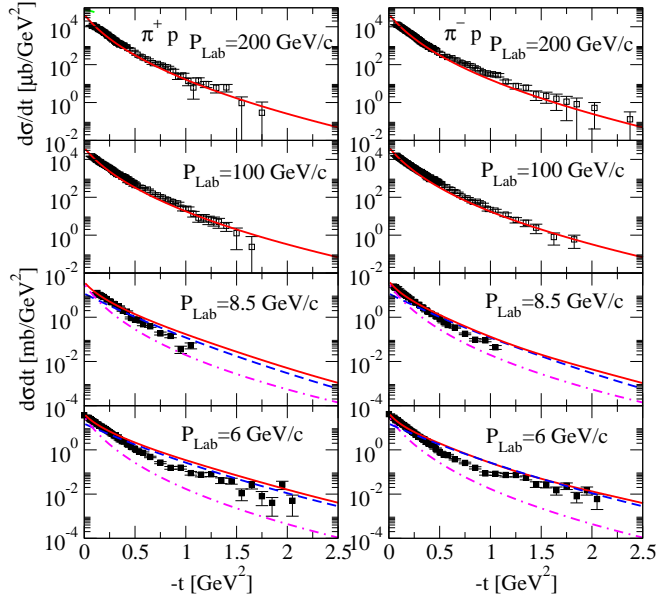}%
\caption{Differential cross sections $d\sigma/dt$ for $\pi^+ p$
(left) and $\pi^- p$ (right) elastic scattering at P$_{\rm Lab}=200$,
100, 8.5, and 6 GeV/c, respectively. Dashed and dash-dotted curves
in the lower two panels are the contributions of $f_2$ and
Pomeron. Data at 100 and 200 GeV/c pion momenta are taken from
Ref. \cite{akerlof} and data at 6.0 8.5 GeV/c are from Refs.
\cite{ambarts,harting}, respectively.} \label{fig7}
\end{figure}

\begin{figure}[]
\centering \epsfig{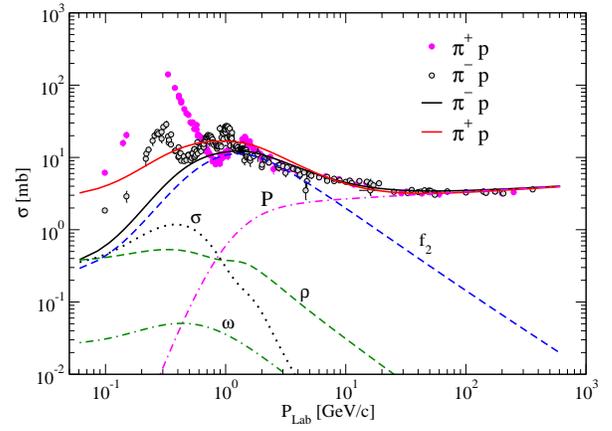}%
\caption{Total cross section $\sigma$ for elastic reactions $\pi^+
p$ (blue) and $\pi^- p$ (red). Notations for $f_2$ and Pomeron are
the same as in Fig. \ref{fig7} for both processes. The difference
between $\pi^+p$ and $\pi^-p$ cross sections is due to the roles
of $\rho+\omega$ exchanges. Dominance of $f_2+$ Pomeron exchanges
are apparent. World data are taken from
PDG \cite{amsler}.} \label{fig8}
\end{figure}

Figure \ref{fig7} presents differential cross sections for
$\pi^+p$ and $\pi^-p$ elastic scatterings. In each reaction those
cross sections at high momenta $P_{\rm Lab}=100$ and 200 GeV/c in
the upper two panels are used to determine the Pomeron trajectory,
while all the physical constants we take for $f_{\pi qq}$,
$\beta_u$, $\beta_d$, and $\mu^2_0$ are fixed as before. Nevertheless,
however, there is no criterion for what value we have to choose
for the parameter $\mu$ at present, because the existing data are
insensitive to a change of $\mu$. In this work we choose $\mu=1$
GeV and $n=2$ for illustration purpose. Then, by leaving the slope
and intercept of the Pomeron trajectory $\alpha_\mathbb{P}(t)$
free parameters to fit to high energy data, we obtain a good
agreement with the energy and $t$-dependence of the cross sections
at the choice of
\begin{eqnarray}\label{pomeron-traj}
\alpha_\mathbb{P}(t)=0.12\,t+1.06
\end{eqnarray}
for  Eq. (\ref{Pomeron-traj}). We note that the slope in the $\pi
N$ scattering is consistent with Ref. \cite{huang}, but by the
factor of 1/2 slower than that of the Pomeron
$\alpha_\mathbb{P}(t)=0.25\,t+1.08$ fitted to total cross section
of $\pi N$ reaction \cite{dl-tot}. Note that
the slope of the total cross section at high energies
given as the energy to the power
$\sim s^{0.0808}$ (thus, $\alpha(0)=1.08$ by $\sigma\simeq s^{\alpha(0)-1}$)
is by far different from that of the elastic cross section of the present
issue as can be seen in Fig. \ref{fig8}.

In Fig. \ref{fig8} we present total elastic cross sections for
$\pi^+p$ and $\pi^-p$ where the contribution of the meson exchange
as well as that of the Pomeron are shown. A few remarks are in order on
the features of meson exchanges;
The vector mesons $\rho$ and $\omega$ are responsible for the
difference between $\pi^+p$ and $\pi^-p$ cross sections, as shown
from threshold up to $P_{\rm Lab}\approx 2$ GeV/c. At high
momenta, the exchanges of $f_2$ and Pomeron in the isoscalar
channel are dominant over $\rho$ and $\omega$ so that the two
cross sections coincide with each other, which should be
distinguished from the difference between the total cross sections
at high energy as in Ref. \cite{mathieu,phillips,dl}. Thus, the
reaction mechanisms of $\pi^\pm p$ elastic reactions are
characterized by the dominance of the natural parity exchange in
the isoscalar channel.

\begin{figure}[]
\centering \epsfig{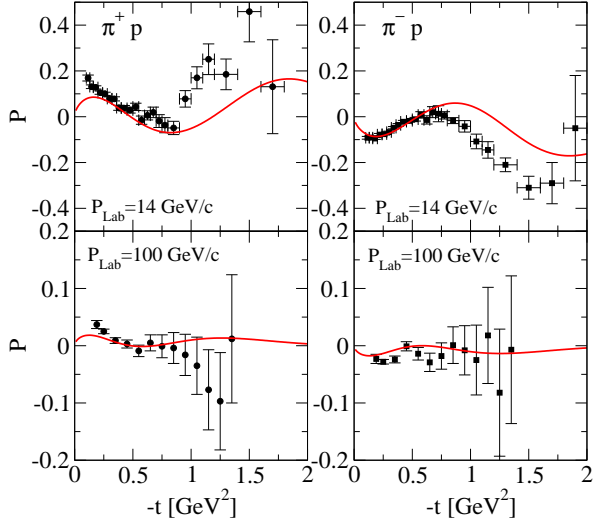}%
\caption{Polarization asymmetry $P(t)$ for $\pi^+ p$ (left) and
$\pi^-p$ (right) elastic scattering at $P_{\rm Lab}=14$ and $100$
GeV/c. Predictions from the model are in good agreement with data.
Data are taken from Refs. \cite{borghini,auer}.} \label{fig9}
\end{figure}

Polarization of target proton  is the observable that could verify
the accuracy of model predictions for the experimental
measurement. To show the validity of the present model we present
the polarization for $\pi^\pm p$ reactions in Fig. \ref{fig9} at
the intermediate and high momenta $P_{\rm Lap}=14$ and 100 GeV/c.
In these results the mirror symmetry between $\pi^+p$ and $\pi^-p$
which is the feature of the polarizations of opposite charges is
well reproduced in any momentum range. In particular, polarizations
are sensitive to the contribution of $f_2$ exchange with the
coupling constant $g^{(2)}_{f_2NN}=0$ for the better agreement
with data.

\section{baryon resonances below $W\leq$ 2 GeV}

In this section we present the nucleon resonances in the
energy-dependence of cross section based on the $t$-channel
exchanges as discussed in previous sections. More data from the
angular distributions and spin polarizations could make improved
the resonance parameters more precisely. The most updated analysis
for the nucleon resonances can be found in the SAID programme with
Ref. \cite{workman}.
However, such a fine-tuning is beyond the scope
of the present work and our aim here is to demonstrate how the
Regge poles are well suited for the nucleon resonance of
the Breit-Wigner form in the reaction amplitude,
\begin{eqnarray}\label{total}
{\cal M}=\left({\cal M}_{Regge}+{\cal M}_\mathbb{P}\right)+{\cal
M}_R\,.
\end{eqnarray}

\begin{table}[t]
\caption{$\Delta$ and $N^*$ resonances in $\pi N$ scatterings.
Mass and width in unit of MeV are taken at the Breit-Wigner fit in
the PDG. Process I stands for $\pi^-p\to\pi^0n$, II for
$\pi^-p\to\pi^-p$, and III for $\pi^+p\to\pi^+p$, respectively.
$^{(*)}$In III, in addition to the Gaussian damping factor with
the parameter $d$, the cutoff function in Eq. (\ref{pionff}) with
$n=1$ and $\mu=m_\pi$ is applied to $\Delta^{++}$ multipole. }
\begin{tabular}{c|c|c|c|c|c}
\hline
Process&Resonance                & $M_R$ & $\Gamma_R$ & $c_R$ & $d$  \\
\hline\hline
       &$\Delta^0(1232)\ P_{33}$& $1232$ & $125$ & $0.5$ & $0.3$  \\%
I      &$N^*(1440)\ P_{11}$     & $1440$ & $400$ & $0.6$ & $0.7$  \\%
       &$N^*(1535)\ S_{11}$     & $1510$ & $150$ & $0.4$ & $0.5$  \\%
       &$N^*(1650)\ S_{11}$     & $1650$ & $125$ & $0.7$ & $0.4$  \\%
       &$N^*(1720)\ P_{13}$     & $1720$ & $250$ & $0.3$ & $0.4$  \\%
       &$\Delta^0(1905)\ F_{35}$& $1900$ & $300$ & $0.2$ & $0.1$  \\%
\hline
&$\Delta^0(1232)\ P_{33}$ & $1232$ & $125$ & $0.35$ & $0.4$  \\%
II&$N^*(1440)\ P_{11}$  & $1420$ & $400$ & $-0.5$ & $2.2$  \\%
&$N^*(1535)\ S_{11}$  & $1510$ & $150$ & $0.6$ & $0.5$  \\%
&$N^*(1650)\ S_{11}$ & $1650$ & $125$ & $0.75$ & $0.4$  \\%
&$N^*(1720)\ P_{13} $ & $1720$ & $250$ & $0.3$ & $0.4$  \\%
&$\Delta^0(1905)\ F_{35}$ & $1900$ & $300$ & $0.2$ & $1$  \\%
\hline
III&$\Delta^{++}(1232)\ P_{33}$ & $1235$ & $120$ & $2$ & $0.03^{(*)}$  \\%
&$\Delta^{++}(1905)\ F_{35}$ & $1900$ & $400$ & $0.5$ & $0.9$  \\%
\hline  \hline
\end{tabular}\label{tb2}
\end{table}

By the conventional definition of the nonrelativistic scattering
amplitude as in the Appendix A, we write the scattering amplitude
as
\begin{eqnarray}
{\cal M}_R={8\pi W\over\sqrt{4MM'}}\sqrt{{k\over q}}\,
\left[F(s,\theta)+i\sigma\cdot \hat{n}\,
G(s,\theta)\right],
\end{eqnarray}
with $\hat{n}=\hat{k}\times\hat{q}/\sin\theta$, and
consider the spin non-flip and flip amplitudes
to be of the form, i.e.,
\cite{lennox}
\begin{eqnarray}
&& F(s,\theta)={1\over k} \sum_R {c_R \left(J_R+1/2\right)\over
\epsilon_R-i}e^{-d\epsilon_R^2}P_l(\cos\theta)\,,\label{bw1}\\
&& G(s,\theta)={1\over k} \sum_R
{c_R\left(-1\right)^{J_R-l+1/2}\over
\epsilon_R-i}e^{-d\epsilon_R^2}{dP_l(\cos\theta)\over
d\cos\theta}\,,\label{bw2} \ \ \ \ \
\end{eqnarray}
with the $d$ in the Gaussian type of the damping factor to adjust
the width of the resonance. Here, $c_R=I_R X_R$ is a sort of the
coupling strength of the resonance $R$ originating from the
product of the Clebsch-Gordon coefficient for isospin and
elasticity. $\epsilon_R=(M_R^2-s)/M_R\Gamma_R$ is the $s$-channel
pole with the mass and full width of the resonance $R$. $k$ and
$\theta$ are the momentum and scattering angle in the c.m. system.
$J_R$ is the spin of the resonance.

\begin{figure}[h]
\centering \epsfig{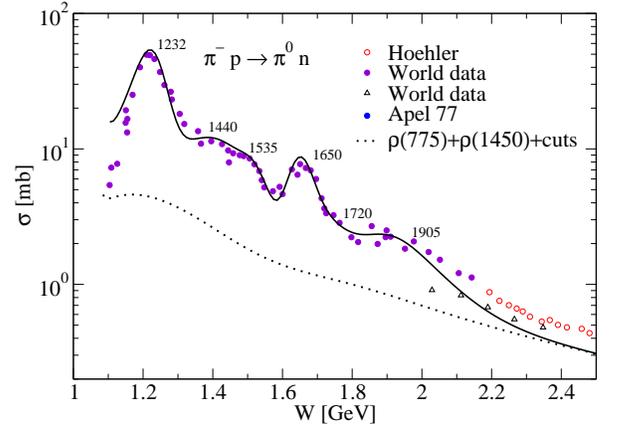}%
\caption{Nucleon resonances in $\pi^- p\to\pi^0n$ reaction.}
\label{fig10}
\end{figure}

\begin{figure}[h]
\centering \epsfig{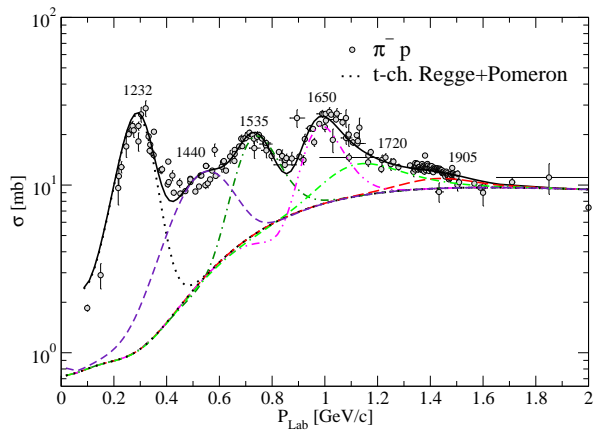}%
\caption{Nucleon resonances in $\pi^- p\to\pi^- p$ reaction. The
respective contributions of nucleon resonances are presented.}
\label{fig11}
\end{figure}

\begin{figure}[hb]
\centering \epsfig{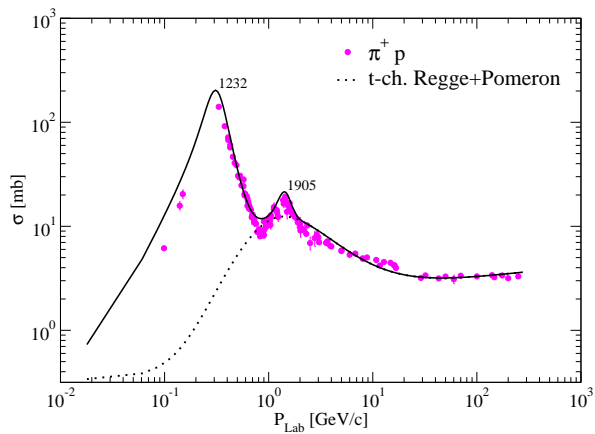}%
\caption{Nucleon resonances in $\pi^+ p\to\pi^+ p$ reaction.
Resonances $\Delta(1232)+\Delta(1905)$ reproduce the peaks.}
\label{fig12}
\end{figure}

Figure \ref{fig10} shows the total cross section for
$\pi^-p\to\pi^-n$ in which case the $t$-channel meson exchanges in
Eq. (\ref{full}) constitutes a background contribution upon which
nucleon resonances are mounting. Nucleon resonances
$\Delta(1232)$, $N^*(1440)$, $N^*(1535)$, $N^*(1650)$,
$N^*(1720)$, and $\Delta(1905)$ are introduced with their
parameters fitted to the total cross section data as in Table
\ref{tb2}. Note that, unlike the Ref. \cite{dolen}, the
contribution of the $t$-channel exchanges in the present
calculation are not passing through the average of the cross
section on the energy interval below $W\approx 2$ GeV, and we
expect that the problem of double-counting should be
insignificant.

The resonance structures in $\pi^-p\to\pi^-p$, and
$\pi^+p\to\pi^+p$ reactions are presented in Figs. \ref{fig11},
and \ref{fig12}. It is worth remarking that the scalar meson
coupling constant $f_{\sigma\pi\pi}=-0.5$ is taken with its sign
reversed, because it is advantageous to alleviate the problem of
double-counting by reducing the contribution of the $\sigma$ meson
which may overlap with resonance. On the other hand, we have to
neglect the threshold divergence of the $\Delta$ pole in Figs.
\ref{fig10}, and \ref{fig11} which are not covered up by the
Gaussian damping factor in the multipoles. In practice, it is the
drawback of the present model calculation of the resonances,
formulated as in Eqs. (\ref{bw1}) and (\ref{bw2}). Moreover, in
the case of $\pi^+p\to\pi^+p$ reaction  where we have to reproduce
the $\Delta^{++}(1232)$ pole with such a wide width that amounts
to 500$\sim$ 600 MeV as can be seen in Fig. \ref{fig12}, the
threshold divergence is even worse. In order to suppress the
strong divergence near threshold, we apply the cutoff function in
Eq. (\ref{pionff}) for the $\Delta^{++}(1232)$ pole with $n=1$ and
$\mu=m_\pi$ in Eq. (\ref{lambdamass}), in addition to the Gaussian
damping factor.

\section{summary and conclusions}

In this work we have investigated $\pi^-p\to\pi^0n$ charge
exchange and $\pi^\pm p\to\pi^\pm p$ elastic reactions up to
incident pion momentum $P_{\rm Lab}\approx 250$ GeV/c to provide a
theoretical framework that could validate a consistency of
the coupling strengths and forms of interaction Lagrangians between
hadrons at low and high energies.
For a description of the reaction at the Regge realm
we utilizes  the relativistic Born amplitude for the reggeization
of the $t$-channel meson exchange. Through the reproduction
of reaction cross sections the reaction mechanism
by the $t$-channel meson and Pomeron exchanges
are analyzed with the coupling constants for hadron interactions
sharing with other hadron reactions, e.g., photoproductions
of vector mesons.

A unique role of vector meson $\rho(775)$ in the charge exchange
reaction is investigated.
Given the single $\rho$ exchange with the deep dip at the
NWSZ point $-t=0.51$ GeV$^2$, the dip-filling mechanism
for the differential cross section needs $\rho$-$f_2$ and
$\rho$-Pomeron cuts. In order
to reproduce the spin polarization a second
$\rho(1450)$ Regge-pole is called for with the trajectory
predicted from the
relativistic quark model, though
the coupling constants of the $\rho(1450)$ are treated as
free parameters. These theoretical entities yield the Regge
description of the charge exchange process to a good degree.

The exchange of a soft Pomeron is newly constructed
from the quark-Pomeron coupling
picture and applied successfully for $\pi^\pm p$ elastic
scatterings with the trajectory quite  different from
that from the total cross sections for $\pi N$ reaction.
The difference between
the $\pi^+p$ and $\pi^-p$ elastic cross sections is insignificant
because of the minor roles of $\rho+\omega$ exchanges,
while $f_2 (1275)$ and Pomeron exchanges
are dominant in the over all range of pion momentum.
Polarizations are well reproduced with the mirror symmetry
reflected between the two reactions of opposite charges.

Nucleon resonances below $W\leq2$ GeV are reproduced in three channels,
$\pi^-p\to\pi^0n$, $\pi^+ p\to\pi^+p$, and $\pi^-p\to\pi^-p$,
and they are consistent with existing data within the masses, widths,
and branching fractions reported in the PDG.
These findings illustrate how the $t$-channel Regge poles
in the present framework do well for the analysis of nucleon resonances
in the low energy region as well as the description of the reactions
at high energies.

       \section*{Acknowledgments}
This work was supported by the National Research Foundation of
Korea grant (Grant No. NRF-2017R1A2B4010117), and partially funded
by (Grant No. NRF-2016K1A3A7A09005580).
\\

\appendix
\section{Partial wave expansion for nucleon resonance}

The scattering amplitude in the $\pi N$ c.m. system is defined by
\begin{eqnarray}
{\sqrt{4MM'}\over8\pi W}{\cal
M}=\chi^\dag\left[F(s,\theta)+i\sigma\cdot \hat{n}\,
G(s,\theta)\right]\chi
\end{eqnarray}
with our convention for the normalization constant
$N=\sqrt{E+M\over2M}$ for the Dirac spinor.  Here $\chi$ is the
$2\times 1$ pauli spinor with spin and isospin indices understood.

The differential cross section is calculated by the equation
\begin{eqnarray}
{d\sigma\over d\Omega}={q\over k}\left|{\sqrt{4MM'}\over8\pi
W}{\cal M}\right|^2= \left|F\right|^2+\sin^2\theta \left|G\right|^2\,.
\end{eqnarray}
The spin non-flip and flip parts of the scattering amplitude are
expanded with the orbital momentum $l$ and the total angula
momentum $J$,
\begin{eqnarray}
&& F(s,\theta)=\sum_{l=0}
\left[\left(l+1\right)f_{l+}(s)+lf_{l-}(s)\right]P_l(\cos\theta)\,,\label{bw11}\\
&&
G(s,\theta)=\sum_{l=1}\left[f_{l+}(s)-f_{l-}(s)\right]
{dP_l(\cos\theta)\over d\cos\theta}. \label{bw22} \ \ \ \ \
\end{eqnarray}
Each partial wave of $\alpha(=l\pm)\,$ is related to the phase shift
by
\begin{eqnarray}
f_{\alpha}(s)={1\over 2ik}\left(e^{2i\delta_{\alpha}}-1\right).
\end{eqnarray}
The energy-dependence of the partial wave for the spin non-flip
and flip amplitudes in Eqs. (\ref{bw11}) and (\ref{bw22})
are parameterized as  in
Eqs. (\ref{bw1}) and (\ref{bw2}).

\end{document}